%

\magnification 1200
\baselineskip=17pt


\centerline{\bf FERMION ANALOGY FOR LAYERED SUPERCONDUCTING}
\bigskip
\centerline{\bf FILMS IN PARALLEL MAGNETIC FIELD}
\vskip 50pt
\centerline{J. P. Rodriguez}
\medskip
\centerline{\it Dept. of Physics and Astronomy,
California State University,
Los Angeles, CA 90032, USA}
\centerline{{\it and Instituto de Ciencia de Materiales,
Consejo Superior de Investigaciones Cientificas,}}
\centerline{{\it Universidad Autonoma de Madrid,
Cantoblanco, 28049 Madrid, Spain.}\footnote*{Present address.}}
\vskip 30pt
\centerline  {\bf  Abstract}
\vskip 8pt\noindent
The equivalence between the Lawrence-Doniach model for films of
extreme type-II layered superconductors and a generalization of
the back-scattering model for spin-1/2 electrons in one dimension
is demonstrated.  This fermion analogy is then exploited to obtain
an anomalous $H_{\parallel}^{-1}$ tail 
for  the parallel equilibrium magnetization of the minimal
double layer case in  the limit of high
parallel magnetic fields $H_{\parallel}$ for temperatures
in the critical regime.

\bigskip
\noindent
PACS Indices: 74.20.De, 74.20.Mn, 74.60.-w
\vfill\eject
The advent of high-temperature superconductors has re-invigorated
the study of layered superconductivity,$^{1-3}$  wherein adjacent
layers are Josephson coupled.  An issue that remains open is the
question of whether or not layer decoupling occurs in the presence of
a parallel magnetic field.$^{4-9}$  The standard  phenomenological model 
used to study this problem theoretically is given by the 
Lawrence-Doniach (LD) free-energy functional, which in the absence
of fluctuations of the magnetic field reads$^9$
$$\eqalignno{
E_{\rm LD} =
J_{\parallel}\Biggl[
\sum_{l = 1}^{N}\sum_{\vec r}
{1\over 2}(\vec\nabla\theta)^2
+\gamma_*^{\prime -2} 
\sum_{l = 1}^{N-1}\sum_{\vec r}\{1-{\rm cos}[\theta(\vec
r,l+1)-
\theta (\vec r,l)-A_z(\vec r, l)]\}\Biggr].\cr
&&(1)\cr}$$
Here, $\theta (\vec r, l)$ denotes the phase of the superconducting
order parameter, which lives on $N$ equally spaced square lattices
with respective coordinates $l$ and  $\vec r = (x,y)$,
while the parallel magnetic field $B_{\parallel} = (\Phi_0/2\pi d)
b_{\parallel}$ directed along the $y$ axis 
is related to the vector potential above by
$A_z = -b_{\parallel} x$, where $d$ represents the spacing between 
layers.  Note that $\vec\nabla = (\Delta_x, \Delta_y)$, with
$\Delta_{\mu}\theta (r) = 
\theta (r + \hat\mu) - \theta (r)$.
Also, $J_{\parallel}$ is a measure of the in-plane phase rigidity,
while $\gamma_*^{\prime}$ is directly related to the mass anisotropy
parameter $\gamma = (m_{\perp}/m_{\parallel})^{1/2}$ of the superfluid
motion within ($m_{\parallel}$) and between ($m_{\perp}$) planes.  
For the bulk case, $N\rightarrow\infty$, where magnetic screening
effects must be included, it was first claimed by Efetov$^4$ that layers
decouple in the mixed phase for parallel fields $B_{\parallel}$
above the characteristic scale 
$B_*^{\parallel}\sim \Phi_0/\gamma d^2$.  Recent work, however,
finds no evidence for a true phase transition as a function of
field.$^{6-9}$

The author has recently studied the relatively simpler problem of
layered thin films of extreme type-II superconductors
($\lambda_L\rightarrow\infty$) in parallel magnetic field.$^9$
In this limit,
magnetic screening effects are negligible and the problem reduces
to the study of $N$ weakly coupled square-lattice $XY$ models
in the presence of uniform frustration; i.e., the thermodynamics
is determined by the energy functional 
$E_{XY} = -\sum_{r,\mu} J_{\mu} {\rm cos}
[\Delta_{\mu}\phi(r) - A_{\mu}(r)]$, where $J_x = J_{\parallel}
= J_y$ and $J_z = J_{\parallel}/\gamma^{\prime 2}$,
with   $\gamma^{\prime}=\gamma d/a$, and
where $A_{\mu} = (0, b_{\perp} x, -b_{\parallel}x)$.  Here
 $a$ denotes the square lattice constant.  
For thin enough films, $Nd\ll\gamma^{\prime} d\ll\lambda_L$, in 
perpendicular magnetic fields,$^{10}$
$B_{\perp} = (\Phi_0/2\pi a) b_{\perp}
\gg\Phi_0/\gamma^2 d^2$, the thermodynamics of this model
factorizes into parallel and perpendicular pieces that
correspond  respectively to Josephson vortices in between
layers and to 2D  perpendicular vortices within each layer.$^{9}$  
The latter is corroborated by
Monte-Carlo simulation.$^{11}$  Physically, this factorization is
due to the fact that the modified Josephson penetration length 
$\gamma^{\prime} d$
sets the minimum perpendicular size for vortex loops that traverse
many layers.$^{12}$  The present factorization prevails in the
presence of both  parallel and perpendicular magnetic field 
as well, with  the parallel 
thermodynamics determined
by the following partition function for a layered Coulomb gas 
ensemble:$^{9,13}$
$$\eqalignno{
Z_{\rm CG} =  \sum_{\{ n_z(\vec r, l)\}} {\rm exp}
\Biggl\{ -{1\over{2\beta_{\parallel}}}\sum_{l=1}^{N}
\sum_{\vec r, \vec r\,^{\prime}}[&n_z(\vec r, l-1) - n_z(\vec r, l)]
G^{(2)}(\vec r - \vec r \,^{\prime})
[n_z(\vec r\,^{\prime}, l-1) - n_z(\vec r\,^{\prime}, l)]\cr
&-i\sum_{l=1}^{N-1}\sum_{\vec r} n_z(\vec r, l) A_z(\vec r, l)
-{1\over{2\beta_{\perp}}}\sum_{l=1}^{N-1}
\sum_{\vec r} n_z^2(\vec r, l)\Biggr\},& (2) \cr}$$
where $n_z(\vec r, l)$ is an integer field over the layered structure
that describes inter-layer fluxon excitations,$^{14}$
with the fields at the boundary layers set to
$n_z(\vec r, 0) = 0 = n_z(\vec r, N)$, and where
$G^{(2)} = -\nabla^{-2}$ is the Greens function for the square
lattice.
Here, $\beta_{\parallel} = J_{\parallel}/k_B T$, while
$\beta_{\perp} = \beta_{\parallel}/\gamma^{\prime 2}$.
(Note that the assumption that magnetic screening be absent
requires that the perpendicular component of the magnetic field
exceed $\Phi_0/\lambda_L^2$.)
By extending Polyakov's Hubbard-Stratonovich
transformation of the neutral Coulomb gas in the
plasma phase,$^{15}$ it can further be shown$^9$ that
this layered Coulomb gas is equivalent to the LD model (1) at 
temperatures below the zero-field  decoupling transition
temperature, $T_* = 4\pi J_{\parallel}$, 
in the limit of small fugacity,
$y_0 = {\rm exp}( - \gamma^{\prime 2}/2\beta_{\parallel})$, 
where the effective anisotropy parameter is given by
$\gamma_*^{\prime} = (\beta_{\parallel}/2 y_0)^{1/2}$.

In this paper, we will first establish that the above layered
Coulomb gas ensemble (2) is equivalent to a one-dimensional
(1D) fermion analogy consisting of coupled
chains.  The latter is a generalization of the 
{\it repulsive} back-scattering model for spin-1/2 fermions introduced 
by Luther and Emery (LE),$^{16}$
where the spin is identified with the layer index.  The Hamiltonian
for this model is divided into two parts,
$H = H_{\parallel} + H_{\perp}$, with  
$$\eqalignno{
H_{\parallel} = & \sum_{l = 1}^N\sum_k\Biggl\{
v_F k\Bigl[a^{\dag}(k,l) a(k,l) - b^{\dag}(k,l) b(k,l)\Bigr]
-\mu_l\Bigl[a^{\dag}(k,l) a(k,l) + b^{\dag}(k,l) b(k,l)\Bigr]\Biggr\}\cr
& + U_{\parallel}\sum_{l=1}^N\int dx \Psi_L^{\dag}(x,l)\Psi_R^{\dag}(x,l)
\Psi_L(x,l)\Psi_R(x,l) & (3a)\cr}$$
and
$$\eqalignno{
H_{\perp} = & U_{\perp} \sum_{l=1}^{N-1}\int dx
\Bigl[\Psi_L^{\dag}(x,l)\Psi_R^{\dag}(x,l+1)\Psi_L(x,l+1)\Psi_R(x,l)
+ {\rm H.c.}\Bigr], & (3b)\cr}$$
and 
with field operators       $\Psi_R(x,l) = L_x^{-1/2}\sum_k e^{ikx} a(k,l)$ and
$\Psi_L(x,l) = L_x^{-1/2}\sum_k e^{ikx} b(k,l)$ for right ($R$) and left
($L$) moving fermions.  
We thus have $N$ Tomonaga-Luttinger chains, 
with adjacent chains
coupled via a LE-type repulsive back-scattering interaction 
$(U_{\perp} > 0$).  The 
magnetic flux between consecutive layers in the LD
model (1) is given by $b_{\parallel} = 2\pi(N_{l+1}-N_l)/L_x$, where
$N_l$ denotes the number of spin-less fermions in the $l^{\rm th}$
chain.
The above fermion analogy thus completes
a   triad of equivalent descriptions (1), (2), and (3a,b)  that   
generalize   the known equivalences between the sine-Gordon model,
the 2D Coulomb gas, and the massive Thirring/LE models to $N$
layers.$^{14-18}$   By analyzing the double-layer case ($N=2$) that
corresponds to the original spin-1/2 back-scattering model,$^{16}$   
we find that the equilibrium magnetization for parallel fields shows
an anomalous $B_{\parallel}^{-1}$ tail in the high-field 
limit at temperatures near $T_*$ (see Fig. 1), as opposed to the 
$B_{\parallel}^{-3}$ tail expected from Ginzburg-Landau theory.$^{5,9}$
This is a result of the entropic presure between neighboring
Josephson vortices, which the fermion analogy correctly accounts for.
Nevertheless, we continue to obtain at best only a cross-over behavior as
a function of parallel magnetic field (see Fig. 1), 
which had been claimed earlier 
on the basis of a    semi-classical analysis.$^9$

{\it Equivalence.}  We now proceed to show        that the above 1D
fermion back-scattering model is equivalent to the layered Coulomb
gas ensemble (2) by extending the 
demonstration given by Chui and
Lee$^{17}$ for the minimal case $N=2$.  The $S$ matrix, which measures the
overlap of the unperturbed ($U_{\perp} = 0$)
groundstate $\Phi_0$ with the exact one $\Psi_0$, is
$\langle\Phi_0|\Psi_0\rangle = 
\langle\Phi_0|{\rm exp}(-i\bar T_0
H_{\perp})|\Phi_0\rangle$.
After making the standard canonical transformation to trivialize
the $U_{\parallel}$ term in (3a) and employing the boson representation
for  the back-scattering interaction,$^{16,19}$ $H_{\perp}$, 
a perturbative
expansion of the former exponential yields
$$\eqalignno{
\langle\Phi_0|\Psi_0\rangle = & \sum_{n=0}^{\infty}
[U_{\perp}/(2\pi\alpha)^2]^{2n}
\int_0^{i\bar T_0} d \tau_{2n} ...  \int_0^{\tau_3} d \tau_2 
\int_0^{\tau_2} d \tau_1   
\Bigl(\Pi_{i} \int_0^{L_x} d x_i\Bigr)\cr
&\times  
\sum_{\{n_z(i)\}}\Biggl\{\Biggl\langle
\Pi_{i} {\rm exp}\{n_z(x_i,\tau_i,l_i) e^{\phi}
[\phi_L(x_i,\tau_i,l_i) + \phi_R(x_i,\tau_i,l_i)]\}\Biggr\rangle_0\cr
&\times \Pi_{i} 
{\rm exp}[-i n_z(x_i, \tau_i, l_i)  
A_z(x_i, \tau_i, l_i)]\Biggr\},
& (4) \cr}$$
where $n_z(x_i,\tau_i,l_i) =\pm 1$ are    inter-layer 
fluxon charge distributions$^{13}$ such that
$\sum_i n_z(x_i,\tau_i,l_i) = 0$,
where $A_z(x, \tau, l) = -2(k_{F,l+1} - k_{F,l}) x$ is a function 
of the Fermi wavenumbers, $k_{F,l}$ corresponding to each chain, $l$,
and where the average $\langle ...\rangle_0$ is over the non-interacting
groundstate.
Above, $\phi_j(x,\tau,l) = \psi_j(x, \tau, l+1) - \psi_j(x, \tau, l)$,
where $\psi_j(x, \tau, l)$ is the time evolution of the local operator
$\psi_j(x, l) = {\rm lim}_{\alpha\rightarrow 0}
2\pi L_x^{-1} \sum_k k^{-1} {\rm exp}
(-{1\over 2} \alpha |k| -ikx) \rho_j(k,l)$
that results from the boson representation, with 
the usual particle-hole operators given by
$\rho_R (k,l) = \sum_q a^{\dag}(q+k,l) a(q,l)$ and 
$\rho_L (k,l) = \sum_q b^{\dag}(q+k,l) b(q,l)$.
Notice that $n_z = 1$ corresponds to the choice of
the step-up term for $H_{\perp}$ in the perturbative expansion,
while $n_z = -1$ corresponds to the step-down term.
Also, we have that ${\rm tanh}\, 2 \phi = U_{\parallel}/2\pi v_F$,
as a  result of   the above-mentioned canonical transformation,$^{16}$
which yields a renormalized Fermi velocity equal to
$v_F^{\prime} = v_F\, {\rm sech} \, 2\phi$.
After separating $\psi_j$ into creation and destruction pieces,
changing variables to $y = v_F^{\prime}\tau$ and
$L_y = v_F^{\prime} i\bar T_0$,
and extending the Chui and Lee procedure$^{17}$ to the present
case (4), we obtain the equality
$\langle\Phi_0|\Psi_0\rangle = Z_{\rm CG}$ along with the following
identifications:
$$\eqalignno{
b_{\parallel} = & 2(k_{F,l+1} - k_{F,l}), & (5)\cr
\beta_{\parallel}^{-1} = & 4\pi e^{2\phi}, & (6)\cr
y_0 = & (2\pi)^{-2} (a/\alpha)^2 (|U_{\perp}|/v_F^{\prime}).
& (7)\cr}$$
Notably, since $N_l = \pi^{-1} k_{F,l} L_x$ gives the number of
spinless fermions in a given chain, we obtain the relationship
$b_{\parallel} = 2\pi (N_{l+1} - N_l)/L_x$ 
for the average magnetic induction in between layers $l$ and $l+1$
announced in the introduction.  Notice that the back-scattering
term ($U_{\perp}$) sets the anisotropy parameter, $\gamma_*^{\prime}$,
in the LD model (1),
whereas the trivial intra-chain interaction ($U_{\parallel}$)
sets the temperature.  Finally, if we approximate the $S$ matrix of the
fermion analogy by
$\langle\Phi_0|\Psi_0\rangle =
{\rm exp}\{-i\bar T_0 [E_F(U_{\perp}) - E_F(0)]\}$, where
$E_F(U_{\perp})$  denotes the energy of the groundstate
$\Psi_0$, then we can identify the free energy $G_s- G_n$
of the LD model in parallel field (1) with the latter via
$$(L_y/v_F^{\prime}) [E_F(U_{\perp}) - E_F(0)]
= (G_s - G_n)/k_B T. \eqno (8)$$
Thermodynamic properties of the layered superconductor are in this
way directly related to ground-state properties of the fermion 
analogy.

{\it Double Layer.}  Consider now the minimal $N=2$ case, where the LD
model (1) reduces to the sine-Gordon model, the ensemble (2) reduces
to the conventional 2D Coulomb gas, and where the fermion analogy (3a,b)
is simply the original LE back-scattering model for spin-1/2 fermions.
In this instance, charge and spin particle-hole operators
$\rho_j^{\prime} (k) = 2^{-1/2} [\rho_j^{\prime}(k,1) +
\rho_j^{\prime}(k,2)]$ and  $\sigma_j^{\prime} (k) = 2^{-1/2}
[\rho_j^{\prime}(k,2) -
\rho_j^{\prime}(k,1)]$  can be defined, which separate out
of the canonically transformed Hamiltonian $H_{\parallel}^{\prime}$.
Furthermore, at the special point $2^{1/2} e^{\phi} = 1$
identified by LE, the spin part becomes equivalent to the
one-body Hamiltonian
$$H_{\sigma}^{\prime} = v_F^{\prime} \sum_k 
k(a_k^{\dag} a_k - b_k^{\dag} b_k) 
+ \Delta_{\sigma}\sum_k (a_k^{\dag} b_k 
+ {\rm H.c.})
- 2^{1/2}\mu \sum_k (a_k^{\dag} a_k + b_k^{\dag} b_k)\eqno (9)$$
for spinless fermions, where the spin particle-hole operators are
given by $\sigma_R^{\prime} (k) = \sum_q a_{k+q}^{\dag} a_q$ and
$\sigma_L^{\prime} (k) = \sum_q b_{k+q}^{\dag} b_q$.  Above,
$\Delta_{\sigma} = U_{\perp}(2\pi\alpha)^{-1}$ is the spin gap, while
$\mu = {1\over 2}(\mu_2 - \mu_1)$ is the external field.
We therefore have energy eigenvalues
$\varepsilon_k = \pm (v_F^{\prime 2}k^2 + \Delta_{\sigma}^2)^{1/2}$ for such
fermions.  Notice
that the procedure followed here is to first obtain the spinless
fermion Hamiltonian in the absence of field, and to then add the
trivial chemical potential shift in the presence of field.  Let us
begin by computing the lower-critical field of the double
layer superconductor, and hence turn off the
external field within the fermion model
($\mu = 0$).   The edges of the double layer contribute
Dirac Eq.-type bound states at zero energy that decay as
$e^{\pm x/\lambda_J}$, where 
$\lambda_J = v_F^{\prime}/|\Delta_{\sigma}|$ is the effective
Josephson penetration length.$^{20}$
Then  Eq. (8) implies that  the line-tension of 
a single Josephson vortex is given by 
$\varepsilon_{\parallel} = k_B T |\Delta_{\sigma}|/v_F^{\prime}
 = 2\pi J_{\parallel}/\lambda_J$, since $|\Delta_{\sigma}|$ gives the
energy cost of adding one fermion/flux quantum  to the system,
and since $k_B T = 2\pi J_{\parallel}$ at this special point
[see Eq. (6)].  The lower-critical field is then simply
$H_{c1}^{\parallel} = 4\pi\varepsilon_{\parallel}/\Phi_0$.
Last, we note that the identification (7) implies that the
bare  Josephson penetration length corresponding to the
LD model (1) is equal to
$\gamma_*^{\prime} a = (\pi v_F^{\prime}/|U_{\perp}|)^{1/2} \alpha$
at the present temperature, while
$\lambda_J = 2\pi (v_F^{\prime}/|U_{\perp}|)\alpha$.
In the limit of weak coupling, $U_{\perp}\rightarrow 0$,
the latter renormalized Josephson penetration length
is then much larger  than the corresponding bare scale.
In particular, the ratio of the parallel lower-critical field to the
bare one set by the LD functional
is equal to  $\gamma_*^{\prime} a/\lambda_J\sim \alpha/a\gamma_*^{\prime}$, 
which is  small if
$\alpha\sim a$.   This effect
is a result of vortex wandering.$^{9,22}$ 

To compute the parallel equilibrium magnetization
deep inside the mixed phase, $B_{\parallel}\gg H_{c1}^{\parallel}$,
we first relate
the external field $\mu$ to the parallel magnetic induction in the
absence of inter-chain/layer coupling ($U_{\perp} = 0$);
i.e., Pauli paramagnetism gives
$b_{\parallel} = 2\pi (N_2 - N_1)/L_x = 2\pi\chi_0\mu$,
where $\chi_0 = 4(2\pi v_F^{\prime})^{-1}$.
Since the magnetization is generally given by
$M_{\parallel} = -{\partial\over{\partial H_{\parallel}}}
[(G_s-G_n)/L_x L_y d]$,
where $H_{\parallel} = 
B_{\parallel} - 4\pi M_{\parallel}
= (\Phi_0/2\pi d) h_{\parallel}$ is the magnetic field, 
we have by Eq. (8) that
$$-4\pi M_{\parallel} = (4\pi/\Phi_0) (2\pi k_B T/v_F^{\prime})
(\partial\mu/\partial h_{\parallel})
{\partial\over{\partial\mu}}\{[E_F(U_{\perp}) - E_F(0)]/L_x\}.$$  
But 
$\partial E_F/\partial\mu = -(N_2-N_1) 
= -L_x\chi_0(\mu^2-{1\over 2}\Delta_{\sigma}^2)^{1/2}$,
where we have used $N_2 - N_1 = 2^{1/2} k_F L_x/\pi$ in conjunction
with $2^{1/2}\mu = \varepsilon_{k_F}$.
If we presume that $\partial\mu/\partial h_{\parallel}$ above
is given by $\partial\mu/\partial b_{\parallel}$
in the absence of inter-layer coupling, 
we obtain
$$-4\pi M_{\parallel} = 2^{-1/2} H_{c1}^{\parallel}
\Biggl[\Biggl(1+{B_{\parallel}^2\over{B_*^2}}\Biggr)^{1/2} -
{B_{\parallel}\over B_*}\Biggr]\eqno (10)$$
for the equilibrium magnetization,$^{21}$
where $B_* = (2^{1/2}/\pi)(\Phi_0/\lambda_J d)$ is the
parallel cross-over field.  
[The ratio of this cross-over scale to the 
bare one $B_0\sim \Phi_0/\gamma_*^{\prime} a d$ set by the LD functional 
is again $B_*/B_0\sim \alpha/a\gamma_*^{\prime}$, and hence small if
$\alpha\sim a$.]
This simple functional form is plotted in Fig. 1.
Notably, (10) predicts an anomalous 
$-4\pi M_{\parallel}/H_{c1}^{\parallel} =
2^{-3/2} B_*/B_{\parallel}$ tail for relatively high fields
$B_{\parallel} > B_*$. 
Straight forward minimization of the LD functional (1) in parallel
magnetic field leads to a much weaker $B_{\parallel}^{-3}$ 
dependence at high
fields,$^{5,9}$ with a more pronounced cross-over
at $B_*$.  This indicates that the entropic pressure generated
by fluctuations of the vortex array is {\it not} negligible
at temperatures in the critical
regime, which is  defined by the relationship $k_B T \sim J_{\parallel}(T)$
(see ref. 9).

What,  however, is the nature of the vortex array itself?
Clearly, the mixed phase corresponds to a gapless
Luttinger liquid in terms of the LE model.  Voit has shown that
algebraic spin-density wave (SDW) correlations in space dominate in
such case for $U_{\perp} > 0$.$^{19}$  Hence, by the equivalence 
demonstrated above, this implies that the array of Josephson vortices
displays algebraic long-range order transverse to the applied magnetic
field.  This physics is accurately captured by the SDW meanfield
theory for (3a,b) characterized by the self-consistency equation
$\chi_l = L_x^{-1}\sum_k\langle a^{\dag}(k,l) b(k,l)\rangle$ and gap 
equation $\Delta_l = U_{\parallel}\chi_l + U_{\perp}
(\chi_{l+1} +\chi_{l-1})$.  A standard analysis of this theory$^{23}$
for the case $N=2$ yields a phase diagram identical
to that derived via renormalization group arguments;$^{17}$
i.e., an SDW phase ($\chi_2 = -\chi_1$)  exists for (repulsive)
$U_{\perp} > U_{\parallel}$.  Notice that this implies a leading
dependence $k_B T_* = 4\pi J_{\parallel} [1 + {1\over 4}(\alpha/a
\gamma_*^{\prime})^2]$ for the decoupling transition temperature as 
a function of the anisotropy parameter by Eqs. (6) and (7).   
If an external field
$|\mu| > |\Delta_l|$  is introduced, however, a unitarity catastrophe
occurs, wherein $|\Delta_l|$ acquires an {\it imaginary}  part.$^{23}$
This we take to be a signal of the  Luttinger 
liquid ($k_F\neq 0$) ``instability''.

$N$ {\it Layers.}  Given that the above meanfield theory works quite
well for the minimal case $N=2$, it should be accurate at least with
respect to the phase diagram for general $N$.  For large
$N$, an SDW phase $\chi_{l +  1} = - \chi_l$ is stable for 
(repulsive) $U_{\perp} > {1\over 2} U_{\parallel}$.$^{23}$
We therefore argue by comparison with the match in the case of
$N = 2$ that an anomalous $B_{\parallel}^{-1}$ tail appears in the
equilibrium parallel magnetization as well at high fields
$B_{\parallel} > B_*$.  And  
that the lower-critical field is given
by the double-layer result as well can be
shown by observing that
flux penetration must begin at the edges, each
of  which reduce to a double layer
at low fields and for weak inter-layer coupling.  
The latter can be seen
via  the Coulomb gas analogy (2),
where it becomes obvious that the effect of neighboring layers
is to produce a weak dielectric renormalization in  the effective
2D Coulomb gas that resides at the edge in the limit of weak inter-layer
coupling.$^{22}$

In summary, we have obtained a fermion analogy (3) for extreme type-II
layered superconductors in parallel magnetic field (1).  In the
minimal double-layer case, we predict an anomalous
$H_{\parallel}^{-1}$ tail for the parallel magnetization in the limit of
high parallel magnetic field, $H_{\parallel}$,
for temperatures in the critical regime.  A mean-field treatment
of the fermion
analogy for the $N$-layer case  (3), however, suggests that this dependence
is generic to extreme thin-film geometries.$^{23}$

The author is indebted to E. Rezayi for discussions.  He also
is grateful for the hospitality of the Theoretical Division at
Los Alamos National Laboratory, where this work originated.
This work
was supported in part by
National
Science Foundation grant DMR-9322427 and the Spanish Ministry for
Science and Education.

\vfill\eject
\centerline{\bf References}
\vskip 16 pt

\item {1.} W.E. Lawrence and S. Doniach, in
{\it Proceedings of the  12$^{\rm th}$
International Conference on  Low Temperature
Physics (Kyoto, 1970)}, edited by E. Kanda
(Keigaku, Tokyo, 1971) p. 361; see also M. Tinkham, Physica C {\bf 235},
3 (1994).

\item {2.} L.N. Bulaevskii, Zh. Eksp. Teor. Fiz. {\bf 64}, 2241
[Sov. Phys. JETP {\bf 37}, 1133 (1973)].

\item {3.}  G. Blatter, M.V. Feigel'man, V.B. Geshkenbein, A.I. Larkin,
and V.M. Vinokur, Rev. Mod. Phys. {\rm 66}, 1125 (1994). 

\item {4.} K.B. Efetov, Zh. Eksp. Teor. Fiz. {\bf 76}, 1781 (1979)
[Sov. Phys. JETP {\bf 49}, 905 (1979)].

\item {5.} L. Bulaevskii and J.R. Clem, Phys. Rev. B {\bf 44},
10234 (1991).

\item {6.} L.V. Mikheev and E.B. Kolomeisky, Phys. Rev. B {\bf 43},
10431 (1991) (Both this ref. and the following one discusses
an analogy for the LD model
in terms of fermions that
exist  in {\it between} consecutive layers,
which is contrary to the present
work where the fermions live  {\it on} each layer).

\item {7.} B. Horovitz, Phys. Rev. Lett. {\bf 72}, 1569 (1994);
Phys. Rev. B {\bf 47}, 5964 (1993).

\item {8.} S.E. Korshunov and A.I. Larkin, Phys. Rev. B {\bf 46},
6395 (1992).

\item {9.} J.P. Rodriguez, ``On the Decoupling of Layered Superconducting
Films in Parallel Magnetic Field'', Los Alamos preprint LA-UR-95-1908
(cond-mat/9604182).

\item {10.} L.I. Glazman and A.E. Koshelev, Phys. Rev. B {\bf 43}, 2835
(1991).

\item {11.} Y.H. Li and S. Teitel, Phys. Rev. B {\bf 47}, 359
(1993); {\it ibid} {\bf 49}, 4136 (1994).

\item {12.} S. Hikami and T. Tsuneto, Prog. Theor. Phys. {\bf 63},
387 (1980); B. Chattopadyay and S. R. Shenoy,
Phys. Rev. Lett. {\bf 72}, 400 (1994).

\item {13.} J.P. Rodriguez, Europhys. Lett. {\bf 31}, 479 (1995).
       
\item {14.}  S.E. Korshunov, Europhys. Lett. {\bf 11}, 757 (1990);
G. Carneiro, Physica C {\bf 183}, 360 (1991).
  
\item {15.} A.M. Polyakov, Nucl. Phys. B{\bf 120}, 429 (1977);
{\it Gauge Fields and Strings} (Harwood, New York, 1987).

\item {16.} A. Luther and V.J. Emery, Phys. Rev. Lett. {\bf 33}, 589
(1974); P.A. Lee, Phys. Rev. Lett. {\bf 34}, 1247 (1975).

\item {17.} S.T. Chui and P.A. Lee, Phys. Rev. Lett. {\bf 35}, 315
(1975).

\item {18.} S. Coleman, Phys. Rev. D {\bf 11}, 2088 (1975).

\item {19.} J. Voit, Rep. Prog. Phys. {\bf 58}, 977 (1995).

\item {20.} M. Fabrizio and A.O. Gogolin, Phys. Rev. B {\bf 51}, 17827
(1995).

\item {21.}  A result similar to Eq. (10) was obtained long ago by
V.L. Pokrovskii and A.L. Talanov, Zh. Eksp. Teor. Fiz. {\bf 78},
269 (1980) [Sov. Phys. JETP {\bf 51}, 134 (1980)], for the case of
2D incommensurate crystals.

\item {22.} J.P. Rodriguez, to be published in Phys. Rev. B
(cond-mat/9604152).

\item {23.} J.P. Rodriguez, unpublished.

\vfill\eject
\centerline{\bf Figure Captions}
\vskip 20pt
\item {Fig. 1.}  
Shown is the 
equilibrium parallel magnetization characteristic
of a double-layer superconductor deep inside
the mixed phase ($B_{\parallel}\gg H_{c1}^{\parallel}$)
at temperatures in the critical regime.  
Notice that the pronounced $B_{\parallel}^{-1}$ tail resulting from
entropic pressure practically destroys the decoupling
cross-over at $B_*$
(compare with ref. 9).

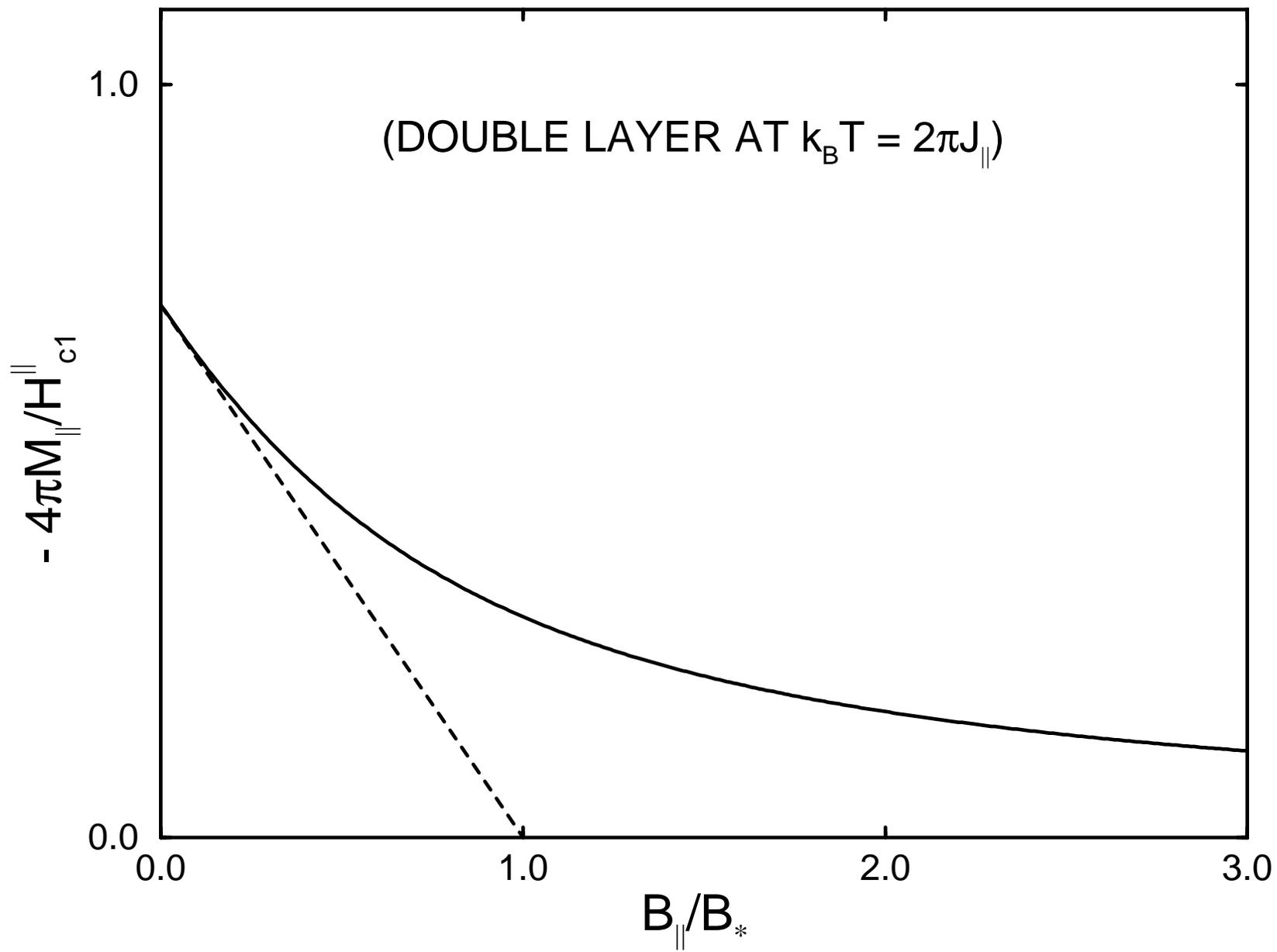
\end